\documentclass{article}
\usepackage{spconf,amsmath,graphicx}
\usepackage{amssymb}
\usepackage{booktabs}
\usepackage{todonotes}

\usepackage{enumitem}
\setlist{nosep, leftmargin=14pt}

\usepackage{mwe} % to get dummy images

\title{K-space Diffusion Model Based MR Reconstruction Method for Simultaneous Multislice Imaging}
\name{Ting Zhao$^{1 2}$ \  Zhuoxu Cui$^{1}$ \ Congcong Liu$^{1}$ \ Xingyang Wu$^{1}$ \ Yihang Zhou$^{1}$ \ Dong Liang$^{1}$ \ Haifeng Wang$^{1}$}

\address{$^{1}$  Shenzhen Institute of Advanced Technology, Chinese Academy of Sciences, Shenzhen, China \\
     $^{2}$ University of Chinese Academy of Sciences, Beijing, China}

\begin{document}

% \todo{Some of the work was supported by the National Natural Science Foundation of China (62271474), the National Key Technology Research and Development Program of China (2023YFB3811403, 2023YFC2411103 and 2023YFF0714201), the International Partnership Program of Chinese Academy of Sciences (321GJHZ2023246GC), the Strategic Priority Research Program of the Chinese Academy of Sciences (XDB0930303), the High-level Talent Program in Pearl River Talent Plan of Guangdong Province (2019QN01Y986), and the Guangdong Basic and Applied Basic Research Foundation (2024A1515012138 and 2023B1515120007).}
% \end{document}
%
\maketitle
\begin{abstract}
 Simultaneous Multi-Slice(SMS) is a magnetic resonance imaging (MRI) technique which excites several slices concurrently using multiband radiofrequency pulses to reduce scanning time. However, due to its variable data structure and difficulty in acquisition, it is challenging to integrate SMS data as training data into deep learning frameworks.This study proposed a novel k-space diffusion model of SMS reconstruction that does not utilize SMS data for training. Instead, it incorporates Slice GRAPPA during the sampling process to reconstruct SMS data from different acquisition modes.Our results demonstrated that this method outperforms traditional SMS reconstruction methods and can achieve higher acceleration factors without in-plane aliasing.\footnote{Ting Zhao and Zhuoxu Cui contributed equally to this work. Some of the work was supported by the National Natural Science Foundation of China (62271474), the National Key Technology Research and Development Program of China (2023YFB3811403, 2023YFC2411103 and 2023YFF0714201), the International Partnership Program of Chinese Academy of Sciences (321GJHZ2023246GC), the Strategic Priority Research Program of the Chinese Academy of Sciences (XDB0930303), the High-level Talent Program in Pearl River Talent Plan of Guangdong Province (2019QN01Y986), and the Guangdong Basic and Applied Basic Research Foundation (2024A1515012138 and 2023B1515120007).}
\end{abstract}

\section{Introduction}
\label{sec:intro}

Simultaneous Multi-Slice(SMS) is a magnetic resonance imaging (MRI) technique, which utilizes multi-band radiofrequency pulses to simultaneously excite multiple slices, thereby significantly reducing the scanning time of MRI, which is especially important for lengthy acquisitions like diffusion tensor imaging(DTI), volumetric T2-weighted scans and time-resolution applications\cite{sms}.
Whereas conventional methods have an unavoidable signal-to-noise ratio(SNR) loss, the SNR in SMS scans increases by the square root of the number of slices. Besides, SMS technique reduces the total scanning time by acquiring data from multiple slices within a single TR, but the sequence parameters remain unchanged, having no impact on blurring or distortion. 
% Additionally, SMS is useful in time-resolution applications like fMRI, where dynamic brain processes can be detected with faster temporal resolution or higher SNR\cite{sms}.

The SMS reconstruction methods include Slice GRAPPA\cite{sg} and SENSE-GRAPPA\cite{sensegrappa}. Based on SENSE GRAPPA, or readout concatenation method (ROC), which creates an extended FOV to separate all the slices, lots of traditional iterative methods and deep learning methods have been proposed like SMS-RAKI\cite{raki} and VCC-RAKI\cite{vccsms}. Compared to ROC methods, Slice-GRAPPA based methods has the ability to quantify the  slice-leakage artifact “L”-factor but only resolve inter-slice aliasing and require other methods to address intra-slice aliasing issues. Currently, there is a lack of deep learning methods that focus on Slice-GRAPPA because it is challenging to integrate SMS data of different MB and CAIPI forms as training data into deep learning frameworks such as unrolling based method and non-unrolling based method.

% Since the introduction of deep learning to the field of MRI reconstruction, a plethora of deep learning methods has emerged. 
% However, due to the variable data structure and difficulty in acquisition of SMS, it is challenging to integrate SMS data as training data into deep learning frameworks like unrolling based method and non-unrolling based method. 

Generative deep learning methods such as score based diffusion model do not require paired data to learn the distribution patterns of MR images, therefor avoiding considering SMS data patterns. Based on our previous Slice-Diffusion method\cite{slicedm}, we further proposed a novel k-sapce diffusion framework combining k-space based diffusion model with Slice GRAPPA which combined a forward process trained by single slice MR images and a reverse process constraint with Slice GRAPPA.
% and have demonstrated superior reconstruction performance in both single-coil and multi-coil MRI data.
% Combining SMS imaging with diffusion model presents itself as a natural idea. Yet, the forward diffusion of multi-shot excitation is challenging to model in the face of complex data structure, and incorporating SMS models into the already time-consuming training and sampling processes of diffusion models significantly increases reconstruction time. 
% Therefore, this study proposed a SMS diffusion model trained with single-slice MRI data but with Slice GRAPPA constraints added during the sampling process. 
The final reconstruction outperforms traditional iterative SMS reconstruction methods and is capable of adapting to higher in-plane acceleration factors.

\begin{figure*}[!t]
    \centering
    \includegraphics[width=1\linewidth]{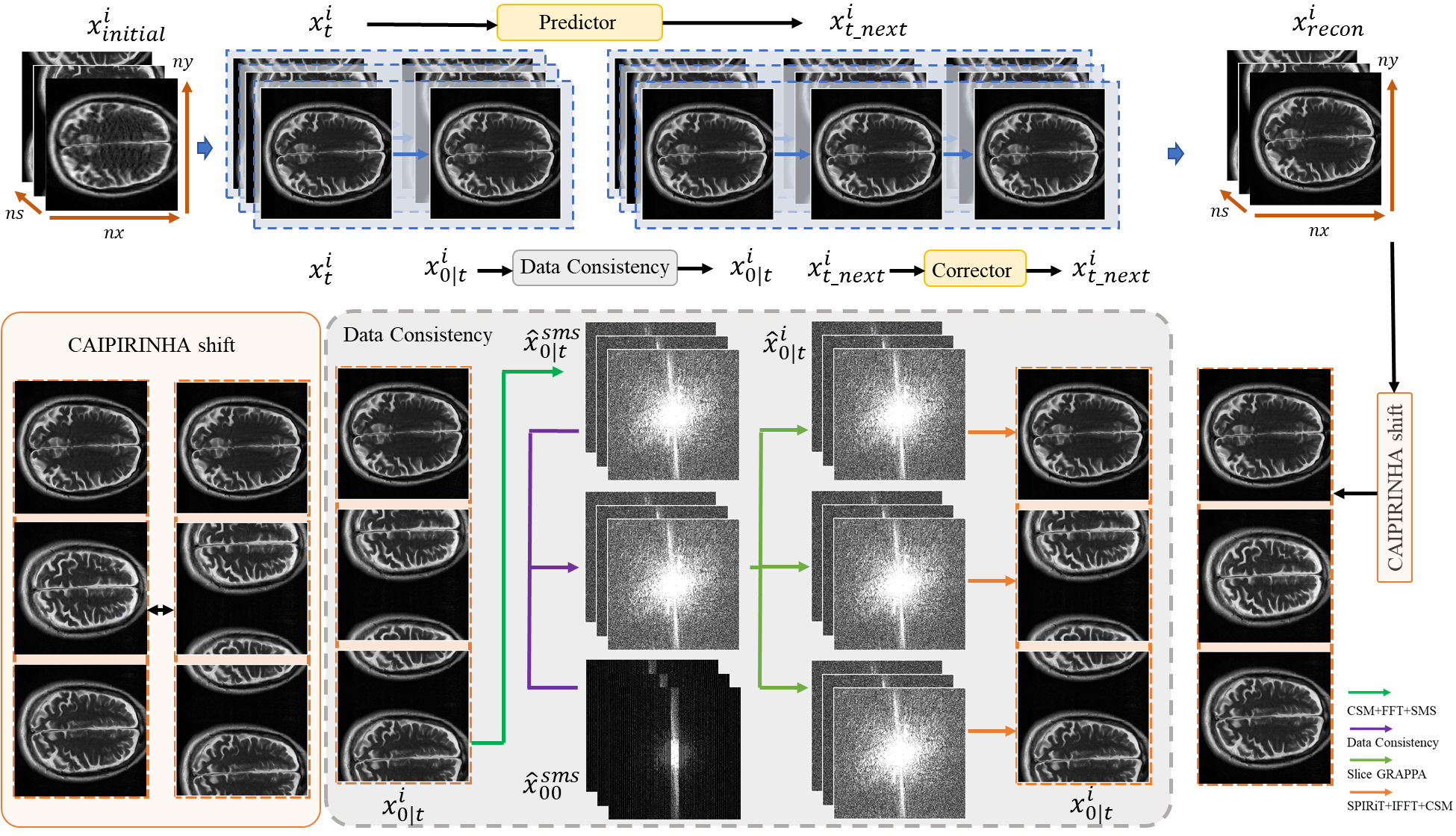}
    \caption{Structure of the SMS Sampling:The SMS sampling begins with the initialized data $x_{initial}^i$, and each slice undergoes a predictor and corrector individually. The SMS constraint is included within the data consistency term of the predictor.
    % According to eq.9, the reverse diffusion process takes place in k-space,but for the convenience of illustration, it is represented as image-domain data. 
    During Data Consistency term, single slice datas $\hat{x}_{0|t}^i$ are combined to get $\hat{x}_{0|t}^{sms}$, and then perform Data Consistency with $\hat{x}_{00}^{sms}$, which follows a Slice GRAPPA to obtain each slice again. At the end of reverse diffusion, the CAIPIRINHA shift is applied to get the final reconstruction.}
    \label{fig:enter-label}
\end{figure*}

\section{METHOD}
\label{sec:format}

\subsection{Slice GRAPPA}
\label{ssec:subhead}

In Slice-GRAPPA, a separate calibration scan is acquired for each slice and used to estimate a set of slice-specific kernels. The kernels are applied to the SMS data to synthesize an entirely new k-space for each slice.
% This procedure is different from GRAPPA, which only synthesizes missing phase encoding lines. Slice-GRAPPA only resolves slice aliasing, so a second parallel imaging step is needed to rectify the in-plane aliasing.
The process of SMS data acquisition can be regarded as:

\begin{equation}
    \hat{x}_{00}^{sms} = D\hat{x}^i+n
\end{equation} where $D$ is the encoding matrix, $n$ is the noise.And Slice GRAPPA kernel $K$ plays its role in this way:

\begin{equation}
    \hat{x}_{alising}^i = K\hat{x}_{00}^{sms}
\end{equation}

In our proposed method, with the acquired data $\hat{x}_{00}^{sms}$, Slice GRAPPA is applied to get initial single-slice data $\hat{x}_{initial}^i$. At each step of the sampling process, we reprocess the single-slice data back into the SMS data, conducting data consistency handling during this procedure, and reapply Slice GRAPPA to separate the data.

%htbp

\subsection{Heat Diffusion for SMS Reconstruction}
\label{ssec:subhead}

% Diffusion models in the image domain have demonstrated remarkable performance in multi-coil MRI data. In addition to these advancements,
In our study, a k-space based diffusion model was used, termed as {\it heat diffusion}\cite{heat}. In Heat Diffusion, the concepts of k-space parallel imaging methods such as GRAPPA and SPIRiT are framed as a process predicting high-frequency missing data based on low-frequency data. 

According to Heat Diffusion, the forward SDE is:
\begin{equation}
    d\hat{z}={\hat{G}}_t\odot \hat{z}(0)dt + \sqrt{\frac{d\sigma^2(t)}{dt}}\overline{S}\overline{S}^*dw
\end{equation}where $G_t$ is a 2-D Gaussian function, $ \sigma(t) $ is the parameter to control the noise level, $\overline{S}$ represents $ \mathbb{F}S\mathbb{F}^{-1}$
, $S$ denotes coil sensitivity. By introducing noise term, we have modeled k-space attenuation as SDE. In contrast to the heat equation, the reverse SDE for equation (3) reads:

\begin{align}
    d\hat{z}=[\hat{G}_t\odot\hat{z}(0)-\frac{d\sigma^2(t)}{dt}\overline{S}\overline{S}^*\nabla_{\hat{z}}logp_t(\hat{z})]dt \nonumber
    \\  +\sqrt{\frac{d\sigma^2(t)}{dt}}\overline{S}\overline{S}^*d\overline{w}
\end{align}Reverse SDE involves an unknown function $\nabla_{\hat{z}}logp_t(\hat{z})$ , the score-matching loss function is reduced to:

\begin{align}
    \theta^* =\underset{\theta}{argmin}\mathbb{E}_t\{\lambda(t)\mathbb{E}_{\hat{z}(t)}\mathbb{E}_{\hat{z}(t)|\hat{z}(0)}[
    ||\overline{S}^* \nonumber \\ (\hat{G}(t)\odot(h_{\theta}(\hat{z}(t),t) - \hat{z}(0)))||_2^2]\}
\end{align} where $h_{\theta}(\hat{z}(t),t)$ means the network.Based on the modeling
of forward and reverse attenuated k-space diffusion, the following
equation:
\begin{align}
    \hat{z}_i=\hat{z}_{i+1}-(\hat{G}_{i+1}-\hat{G}_i)\odot\hat{z}_0+\sqrt{\sigma^2_{i+1}-\sigma^2_i}\overline{S}\overline{S}^*n \nonumber
    \\ +(\sigma^2_{i+1}-\sigma^2_i)\overline{S}\overline{S}^*\nabla_{\hat{z}_{i+1}}logp_{i+1}(\hat{z}_{i+1})
\end{align}enables the reconstruction of missing high-frequency k-space
data using low-frequency ACS data.

\subsection{SMS Sampling}
\label{ssec:subhead}

Due to the distinct multiband (MB) acceleration and complex additional parameters in SMS imaging, such as different CAIPIRINHA shift type, training holistically within the diffusion process is exceedingly intricate. Therefore, single-slice images are utilized for training, enabling the k-space-based diffusion model to learn the overall distribution of MRI images in k-space. During the sampling process, the physical model of SMS is integrated to reconstruct SMS data. The specific reconstruction workflow is depicted in Figure 1:

Before the sampling steps, data initialization is needed. Firstly applying Slice GRAPPA to SMS data to obtain single-slice data with in-plane aliasing. Then, SPIRiT was applied for each slice. Following this initialization, these data were prepared to undergo the reverse diffusion process.

\section{Experiment}
\label{sec:pagestyle}

\subsection{Dataset}
\label{ssec:subhead}

FastMRI Dataset was used to train the Heat Diffusion Model. For sampling test, we simulated SMS data using fastMRI T2 brain dataset and cropped the image to 320 x 320. We set MB=3 with different in-plane acceleration factors. To ensure minimal inter-slice leakage, a 2/3 field-of-view (FOV) CAIPIRINHA shift was applied to the three images, and the central region of k-space was extracted as 32 ACS lines. The corresponding Slice GRAPPA kernel was obtained using the ACS data, along with the SPIRiT kernel for each slice and the coil sensitivity maps. 

% \subsection{Network Architecture and Training}
% \label{ssec:subhead}

% The network structure of attenuated k-space diffusion is the same as that of Heat Diffusion, which directly input multi-coil k-space data to the network. The complex k-space data is split into real and imaginary components and concatenated before input into the network.  The network is trained for 100 epochs in a computing environment using the torch 1.13 library, cuda 11.6 on an NVIDIA A800 Tensor Core GPU.

\subsection{Performance Evaluation}
\label{ssec:subhead}
 For quantitative evaluation, the peak signal-to-noise ratio (PSNR), normalized mean square error (NMSE) value and structural similarity (SSIM)
index were adopted on the image domain.

\section{results and discussion}
\label{sec:typestyle}

\subsection{Compared to traditional SMS methods}
\label{ssec:subhead}

The methodology proposed in this study was compared with traditional Slice-GRAPPA+SNESE and SMS-COOKIE\cite{cookie}. The uniform undersampling mask was used, with in-plane undersampling factors of 3x and 4x. Results of 4x are shown in Fig.2. 
% Under the condition of 3x, it is evident that the images reconstructed by the Slice GRAPPA+SENSE method and SMS-COOKIE exhibit ripple-like artifacts and are inferior in terms of detail when compared to the multi-shot excitation diffusion model we proposed.
Under the condition of 4x, the Slice GRAPPA+SENSE and SMS-COOKIE introduced obvious in-plane aliasing, whereas our proposed method demonstrated better reconstruction results, without in-plane aliasing better high-frequency details. Detailed performance is presented in Table 1.

\begin{figure}[htb]
    \centering
    \includegraphics[width=1\linewidth]{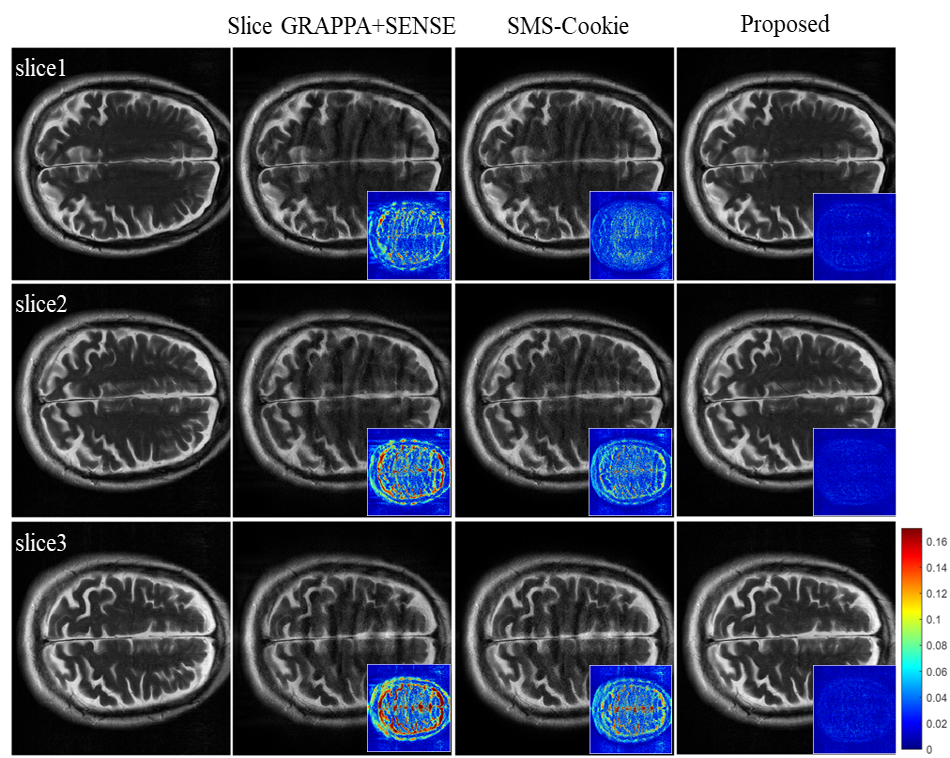}
    \caption{Reconstruction Results of Different Method in 4X}
    \label{fig:enter-label}
\end{figure}

\begin{table}[]
\centering
\begin{tabular}{@{}ccccl@{}}
\toprule

\bottomrule
\end{tabular}
\end{table}

\begin{table}[]
\centering
\begin{tabular}{llll}
\toprule
                & NMSE   & PSNR    & SSIM   \\
\midrule
Slice GRAPPA+SENSE(3x)            & 0.0291 & 30.7509 & 0.8069 \\
SMS-COOKIE(3x)      & 0.0261 & 31.6853 & 0.7446 \\
Proposed(3x) & \textbf{0.0125} & \textbf{38.0801} & \textbf{0.9463} \\
\midrule
Slice GRAPPA+SENSE(4x)            & 0.0614 & 24.4251 & 0.7317 \\
SMS-COOKIE(4x)      & 0.0445 & 27.1895 & 0.6839 \\
Proposed(4x) & \textbf{0.0143} & \textbf{36.9111} & \textbf{0.9356} \\

\bottomrule
\end{tabular}
\caption{Quantitative Comparison for Different Method}
\end{table}

\subsection{Different in plane acceleration factors}
\label{ssec:subhead}

Traditional SMS reconstruction methods are typically limited in in-plane acceleration factors, with noticeable aliasing artifacts emerging at high acceleration, as shown in Fig.2. However, due to the superior reconstruction performance of Heat Diffusion, greater breakthroughs can be achieved in in-plane reconstruction factors. The following figures present the reconstruction results of the model proposed in this study under various in-plane acceleration factors. As illustrated, the method proposed does not exhibit undersampling artifacts even at 8x.  Nevertheless, due to the loss of high-frequency information at higher acceleration factors, some image details gradually become blurred or even disappear as the acceleration factor increases. 

\subsection{Discussion}
\label{ssec:subhead}

We also tried the SMS test data processed by slice GRAPPA using the end-to-end method H-DSLR\cite{hdslr} and the unrolling method MoDL\cite{modl}, but it is challenging for the frameworks of these methods to be compatible with SMS reconstruction based on slice GRAPPA. Therefore, the proposed SMS reconstruction method based on generative diffusion models has significant advantages.

Besides, our current method still has some limitations. At high in-plane acceleration factors, although PSNR and SSIM still maintain relatively high values, the NMSE values indicate that further optimization is needed in terms of high-frequency details. Subsequently, we will add additional constraints to address this issue and compare our method with more methods in multiple dimensions.

\begin{figure}[htb]
    \centering
    \includegraphics[width=1\linewidth]{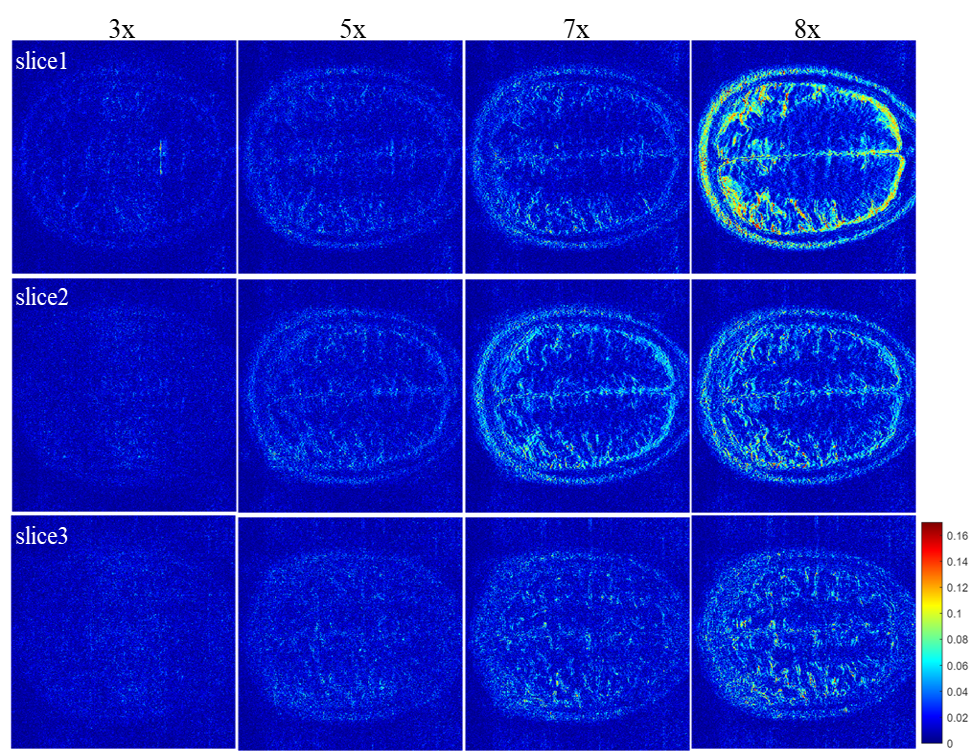}
    \caption{Error Map of Different In-plane acc}
    \label{fig:enter-label}
\end{figure}

\begin{table}[]
\centering
\begin{tabular}{@{}ccccl@{}}
\toprule
     & NMSE   & PSNR    & SSIM   &  \\ 
\midrule
acc3 & 0.0125 & 38.0801 & 0.9463 &  \\
acc4 & 0.0143 & 36.9111 & 0.9356 &  \\
acc5 & 0.0163 & 35.7503 & 0.9285 &  \\
acc6 & 0.0211 & 33.7523 & 0.9182 &  \\
acc7 & 0.0212 & 33.5225 & 0.9123 &  \\
acc8 & 0.0305 & 30.4993 & 0.8954 &  \\ 
\bottomrule
\end{tabular}
\caption{Quantitative Comparison for Different Acc}
\end{table}

\section{Conclusion}
\label{sec:majhead}

This study proposed a SMS reconstruction method based on Heat Diffusion and slice GRAPPA.
% , which leverages single-slice MR images to learn their distribution during the training stage and employed additional physical constraints of SMS during the sampling stage for SMS image reconstruction. 
Our proposed method does not require SMS data for training, thereby avoiding issues associated with the variability and difficulty in obtaining SMS dataset. Moreover, due to the superior reconstruction performance of diffusion model
, our proposed reconstruction method is less likely to exhibit undersampling aliasing artifacts even when using higher in-plane acceleration factors. Compared to traditional SMS reconstruction methods, our method demonstrates superior reconstruction performance.

% \section{Page numbering}
% \label{sec:page}

% Please do {\bf not} paginate your paper.  Page numbers, session numbers, and
% conference identification will be inserted when the paper is included in the
% proceedings.

% \vfill
% \pagebreak

% \section{Copyright forms}
% \label{sec:copyright}

% you submit your paper. We {\bf must} have this form before your paper can be
% published in the proceedings.  

% \section{Acknowledgments}
% \label{sec:acknowledgments}

% Examples of
% appropriate statements include:
% \begin{itemize}
%   \item ``No funding was received for conducting this study. The
%     authors have no relevant financial or non-financial interests to
%     disclose.'' 
%   \item ``This work was supported by […] (Grant numbers) and
%     […]. Author X has served on advisory boards for Company Y.'' 
%   \item ``Author X is partially funded by Y. Author Z is a Founder and
%     Director for Company C.''
% \end{itemize}

\bibliographystyle{IEEEbib}
\bibliography{ref2015}

\end{document}